\documentclass[letterpaper,twocolumn,10pt]{article}
\usepackage{usenix2019_v3}

\usepackage{tikz}
\usepackage{amsmath}
\usepackage{wasysym}

\usepackage[acronym,nohypertypes={acronym},shortcuts]{glossaries}
\usepackage{soul}
\usepackage{xcolor}

\usepackage{multirow}
\usepackage{makecell}
\usepackage{rotating}

\usepackage{caption}
\usepackage{subcaption}

\usepackage[many]{tcolorbox}    	% for COLORED BOXES (tikz and xcolor included)
\newtcolorbox{questionbox}{
    colback=gray!20,
    boxrule=0pt,
    arc=1pt,
    boxsep=2pt,
    left=2pt,
    right=2pt,
    leftrule=3pt,
    colframe=black
}

\newtcolorbox{answerbox}{
    colback=green!95!black!20,
    boxrule=0pt,
    arc=1pt,
    boxsep=2pt,
    left=2pt,
    right=2pt,
    leftrule=3pt,
    colframe=green!75!black
}

\usepackage{url}

\usepackage{breakurl}
\begin{document}
%-------------------------------------------------------------------------------

%don't want date printed
\date{}

\title{\Large \bf SoK: Stealing Cars Since Remote Keyless Entry Introduction and \\ How to Defend From It}

%for single author (just remove % characters)
\author{
{\rm Tommaso Bianchi}\\ % 
University of Padova \\
tommaso.bianchi@phd.unipd.it
\and
{\rm Alessandro Brighente}\\  % 
University of Padova \\
alessandro.brighente@unipd.it
% copy the following lines to add more authors
\and
{\rm Mauro Conti}\\  % 
University of Padova \\ 
Delft University of Technology \\
mauro.conti@unipd.it
\and
{\rm Edoardo Pavan}\\  % 
University of Padova \\
edoardo.pavan.3@studenti.unipd.it
} % end author

\maketitle

%---------------------------------------------------------------------
%%% ACRONYMS
%---------------------------------------------------------------------
\newacronym{rke}{RKE}{Remote Keyless Entry}
\newacronym{rfid}{RFID}{Radio Frequency Identification}
\newacronym{rf}{RF}{Radio-Frequency}
\newacronym{pkes}{PKES}{Passive Keyless Entry and Start}
\newacronym{ask}{ASK}{Amplitude Shift Keying}
\newacronym{fsk}{FSK}{Frequency Shift Keying}
\newacronym{crc}{CRC}{Cyclic Redundancy Checks}
\newacronym{mac}{MAC}{Message Authentication Code}
\newacronym{lf}{LF}{Low-Frequency}
\newacronym{uhf}{UHF}{Ultra-High-Frequency}
\newacronym{rssi}{RSSI}{Received Signal Strength Indication}
\newacronym{prng}{PRNG}{Pseudo-Random Number Generator}
\newacronym{dst}{DST}{Digital Signature Transponder}
\newacronym{nlfsr}{NLFSR}{Non-Linear Feedback Shift Register}
\newacronym{lfsr}{LFSR}{Linear Feedback Shift Register}
\newacronym{iv}{IV}{Initilization Vector}
\newacronym{nfc}{NFC}{Near-Field Communication}
\newacronym{ble}{BLE}{Bluetooth Low Energy}
\newacronym{uwb}{UWB}{Ultra Wideband}
\newacronym{api}{API}{Application Programming Interface}
\newacronym{gd}{GD}{Guess-and-Determine}
\newacronym{mitm}{MITM}{Man-In-The-Middle}
\newacronym{sdr}{SDR}{Software Defined Radio}
\newacronym{bcm}{BCM}{Body Control Module}
\newacronym{hrp}{HRP}{High Rate PHY}
\newacronym{suc}{SUC}{Secret Unknown Cipher}
\newacronym{aes}{AES}{Advanced Encryption Standard}
\newacronym{puf}{PUF}{Physical Unclonable Function}
\newacronym{qkd}{QKD}{Quantum Key Distribution}
\newacronym{toa}{TOA}{Time of Arrival}
\newacronym{dtoa}{DTOA}{Difference Time of Arrival}
\newacronym{gps}{GPS}{Global Positioning System}
\newacronym{lstm}{LSTM}{Long Short-Term Memory}
\newacronym{snr}{SNR}{Signal-to-Noise Ration}
\newacronym{arib}{ARIB}{Association of Radio Industries and Businesses}
\newacronym{oem}{OEM}{Original Equipment Manufacturer}
%-------------------------------------------------------------------------------
\begin{abstract}
%-------------------------------------------------------------------------------
Remote Keyless Entry (RKE) systems have been the target of thieves since their introduction in automotive industry. 
Robberies targeting vehicles and their remote entry systems are booming again without a significant advancement from the industrial sector being able to protect against them.
Researchers and attackers continuously play cat and mouse to implement new methodologies to exploit weaknesses and defense strategies for RKEs. 
In this fragment, different attacks and defenses have been discussed in research and industry without proper bridging.
In this paper, we provide a Systematization Of Knowledge (SOK) on RKE and Passive Keyless Entry and Start (PKES), focusing on their history and current situation, ranging from legacy systems to modern web-based ones. 
We provide insight into vehicle manufacturers' technologies and attacks and defense mechanisms involving them.
To the best of our knowledge, this is the first comprehensive SOK on RKE systems, and we address specific research questions to understand the evolution and security status of such systems.
By identifying the weaknesses RKE still faces, we provide future directions for security researchers and companies to find viable solutions to address old attacks, such as Relay and RollJam, as well as new ones, like API vulnerabilities.
\end{abstract}

%-------------------------------------------------------------------------------
\section{Introduction}
%-------------------------------------------------------------------------------
The \ac{rke} and \ac{pkes} systems are access control mechanisms adopted in cars to open the doors and start the vehicle via a wireless channel.
Specifically for the automotive industry, it dates back to 1982 when Renault used the patent deposited the year before for an infra-red remote controller~\cite{fuego_keyless_2014}. 
Since its introduction, car manufacturers employed resources and research to improve their systems and provide more secure locks, for example, adding the immobilizer, an anti-theft system that requires authentication to start the engine~\cite{immobilizer_patent}.
In the last year, manufacturers have included the newest web and radio technologies in the automotive scenario, changing the paradigm of entering the vehicle. 
Relevant examples include \ac{pkes} that do not require driver interaction, sophisticated keyless entries via smartphone applications~\cite{android_key_car, apple_key_car}, or \ac{nfc} cards acting as keys working on ultra-wideband signals to provide better security and comfort.
% \hl{we never talked about cards before} 
% \hl{application on the smartphone? or what?}
However, these systems also increase the attack surface that thieves can leverage to enter, start, and steal the car~\cite{stealing_cars}.
Additionally, manufacturers' creation of custom closed-source devices and algorithms leads to vulnerabilities that can be discovered and used by malicious users. 
This could be solved without resorting to security by obscurity, having security researchers find vulnerabilities before attackers can, or that attackers may already use for real-world exploitation.
% \hl{<-- plz rephrase in modo un po' piu' formale}.
Nonetheless, this field is still not getting the required attention by companies that see their cars stolen even nowadays~\cite{car_theft_issue}.
In this paper, we provide the first fully comprehensive review of \ac{rke} and \ac{pkes} technologies, attacks against them, and defenses in the automotive industry.
To the best of our knowledge, no available contribution in the literature is providing a survey or SOK on this topic.
We particularly devote our attention to the uncovered problems, analyzing how these systems evolved during the years, and answering the following research questions to provide an insight into the difficulties and problems to overcome to secure these systems.
\begin{questionbox}
\textbf{\textit{Q1. Did the evolution of the \ac{rke} and \ac{pkes} technologies increase their security?}}
\end{questionbox}
We explore the \ac{rke} history regarding technologies and tools to understand the impact of car hacking and if and how additional security measures were included. 
Secondly,
\begin{questionbox}
\textbf{\textit{Q2. How did attack strategies evolve in time?}}
\end{questionbox}
We analyze the attacks found in the wild and in research papers to study whether the methodologies changed due to the new \ac{rke} systems or whether they are still vulnerable to \textit{legacy} attacks.  We particularly emphasize the attacker's needs and capabilities to open and start a car.
We then focus on the defense strategies.
\begin{questionbox}
\textbf{\textit{Q3. Are currently existing defense strategies deployable and effective?}}
\end{questionbox}
We answer this question by checking if the defenses presented in the literature are deployable and usable by the car manufacturer for their systems or whether they are purely research-oriented.
Lastly,
\begin{questionbox}
\textbf{\textit{Q4.  What do we need to develop an effective and secure remote entry system?}}
\end{questionbox}
To answer this question, we identify the open security issues that affect the \ac{rke} and \ac{pkes} technologies comprehensively accounting for attacks to legacy systems and attacks to newer technologies. 
By answering this question, we provide future research directions in this field.
Our work systematizes the different kinds of attacks discovered by researchers and used by thieves to steal cars, exploiting weaknesses in the \ac{rke} systems. 
We analyze attacks performed in the wild, thus that are authentic and feasible for thieves to access and start a vehicle in a real-world scenario
In this way, we show the impact the \ac{rke} systems can have and the need for secure devices.
The final objective is to answer the research questions, providing a helpful understanding of the history and current state of \ac{rke} systems, analyzing how the technology improved and how the weaknesses made it defenseless against attackers.

The contribution of this work can be summarized as follows: 
\begin{itemize}
    \item We extensively systematize and collect attacks and defenses on \ac{rke} systems, with more than 35 attacks since 2005 and 13 defenses specifically tailored for \ac{rke} proposed in the literature. At the time of writing, we are the first to provide a comprehensive review of the history and current state of \ac{rke} and \ac{pkes} systems security research.
    \item We analyze the different technologies involved in these systems, such as the cryptographic functions and the physical layers used in \ac{rke}.
    \item We identify the weaknesses afflicting \ac{rke} and the evolution of attacks and defenses, pinpointing specific threats that persist until now and new attack surface, providing research directions for further exploration of this field. We specifically answer the research questions we posed, explaining how we reached such conclusions.
\end{itemize}

In the following, we describe the methodology in Section~\ref{sec:methodology} and the background on \ac{rke} and \ac{pkes} and their evolution in Section~\ref{sec:background} and ~\ref{sec:implementation}. In Section~\ref{sec:attacks} and Section~\ref{sec:defenses}, we describe and analyze the attacks and the defenses in the vehicle remote and passive entry field. In Section~\ref{sec:takeways}, we answer the research questions, also trying to redirect the researchers interested in this sector to the new area of investigation (Section~\ref{sec:future}). Finally, we discuss the similar work in Section~\ref{sec:related} and the conclusions in Section~\ref{sec:conclusions}.

%-------------------------------------------------------------------------------
\section{Methodology}\label{sec:methodology}
%-------------------------------------------------------------------------------
The systematization study comprehends all the major attacks and defenses described in the four top conferences in cybersecurity, namely IEEE S\&P, USENIX, NDSS, and CCS, with the addition of works with significant impact from the most respectful hacking and academic conferences, such as DEF CON and Black Hat.
We selected works with a high practical impact, especially regarding the attacker side.
Attacks targeting a fixed code radio signal are trivial, and we leave them out to focus only on the most interesting and recent aspect of \ac{rke} systems.
Thus, our study starts with attacks against rolling codes, a security mechanism described in Section~\ref{sec:background}.
The research on attacks specifically tailoring these technologies started in 2005, with the first security analysis of the cryptographic-enabled \ac{rfid} device also used as \ac{rke} system. 
After that, researchers introduced different attacks and defenses in a cat and mouse play, adding new techniques and methods on both sides. 
In this sense, we describe these techniques associated with their corresponding technologies, analyzing the impact and the actuality of such methods.

%-------------------------------------------------------------------------------
\section{Background}\label{sec:background}
%-------------------------------------------------------------------------------
The \acrfull{rke} system patent dates back to 1981 from the inventor Paul Lipschutz~\cite{lipschutz_rke_patent}.
Renault introduced the first \ac{rke} system in 1982 on board the Renault Fuego vehicle. 
From that moment, manufacturers adapted the car entry systems from the traditional physical key to the more technologically advanced key fob, integrating wireless commands to open or close the car's doors with a simple button press.
This introduced an additional attack vector, as pointed out in different works after the rising use of this system~\cite{survey_2011, survey_2012} and their first attacks. 
Also, two outstanding automotive researchers, Chris Valasek and Charlie Miller, especially pointed this out~\cite{valasek_survey_2014}.

\subsection{Remote Keyless Entry}
Initially, the \ac{rke} system used an infra-red technology, as described in the patent by Lipschutz, with all the deriving difficulties and problems, 
such as the need to point the transmitter precisely towards 
the receiver~\cite{breuls_security_2019, fuego_keyless_2014}.
Now, they are based on radio frequency, with a short-range radio transmitter that communicates with the car in a range of 10-100 meters.
The \ac{rke} system generally comprises a mechanical locking with the physical key blade, the remote entry system itself, and an \ac{rfid}-enabled immobilizer~\cite{immobilizer_patent} to authorize and start the engine once inside the vehicle~\cite{oswald_keynote}. 
The key blade is included to insert and rotate it to start the engine as with a regular, physical key. 
Still, it is not needed to enter the vehicle (except in exceptional cases, such as dead battery or malfunctioning).
The \ac{rke} system usually works using \ac{ask} modulation for the physical wireless signal. 
Smaller groups of devices work on \ac{fsk} modulation. 
The frequency depends on the regulation, standards, and regions, such as 315 MHz and 433.94 MHz for North America and Japan, where they follow the FCC 15~\cite{fcc_15} (for RF devices) and the \ac{arib} standards for radio frequencies such as the STD-T67~\cite{std_t67} and the STD-T93~\cite{std_t93}.
Instead, in Europe, they operate at 868 MHz for Europe, following the CE mark directives for safety and regulations from ETSI EN 300 220 or ETSI EN 302 291~\cite{etsi_en_300220,etsien_302291}.
The traditional \ac{rke} transmitter sends a command with a preamble, an identifier, the encrypted data field, and an integrity check field such as a \ac{crc}.
To achieve security against possible thefts, the \ac{rke} system implements integrity protection mechanisms such as \ac{mac} and authentication.
In general, the \ac{oem} follow the international ISO/SAE 21434 standard for security guidelines of these devices~\cite{isosae_21434}, where specific aspects and requirements for \ac{rke} systems dictate the design, implementation, and maintenance of a secure system.
Indeed, the specific RF parameters are omitted in this international standard due to their regional dependence. 

\subsection{Passive Keyless Entry and Start}
% \hl{first?}
After \ac{rke}, the car manufacturer introduced the \acrfull{pkes}.
In its first implementation, it is a device with the ability to open the vehicle.
Then, implemented in combination with the remote entry and the immobilizer. 
This technology does not require pressing a button, and the key fob is in continuous low-power listening mode to capture "wake-up" messages by the car's transceiver.
% \hl{with low frequency mi fa pensare a ogni quanto manda il pacchetto, se invece e' la portante allora direi at a}
The car broadcasts this message at a low frequency, generally 125 kHz, and wakes up the key fob to start the challenge-response procedure to authenticate and authorize the action. 
The two transceivers share the same secret key so that the key fob can correctly encrypt the challenge, usually composed by the encryption of a random number presented by the vehicle.
Once the key is located inside the car as verified by \ac{rssi} measurement for proximity, the vehicle also allows the ignition with the button for the engine to start.
In \ac{pkes}, we find the combination of \ac{lf} and \ac{uhf} technologies for the challenge-response mechanism, which could also be minimized to two messages in the case the vehicle also advertises its Car Identifier.
Regarding guidelines for building these systems, the same regulations and standards discussed for \ac{rke} also apply to this technology.

\subsection{Rolling Codes}
\begin{figure}
\begin{center}
\includegraphics[width=\columnwidth]{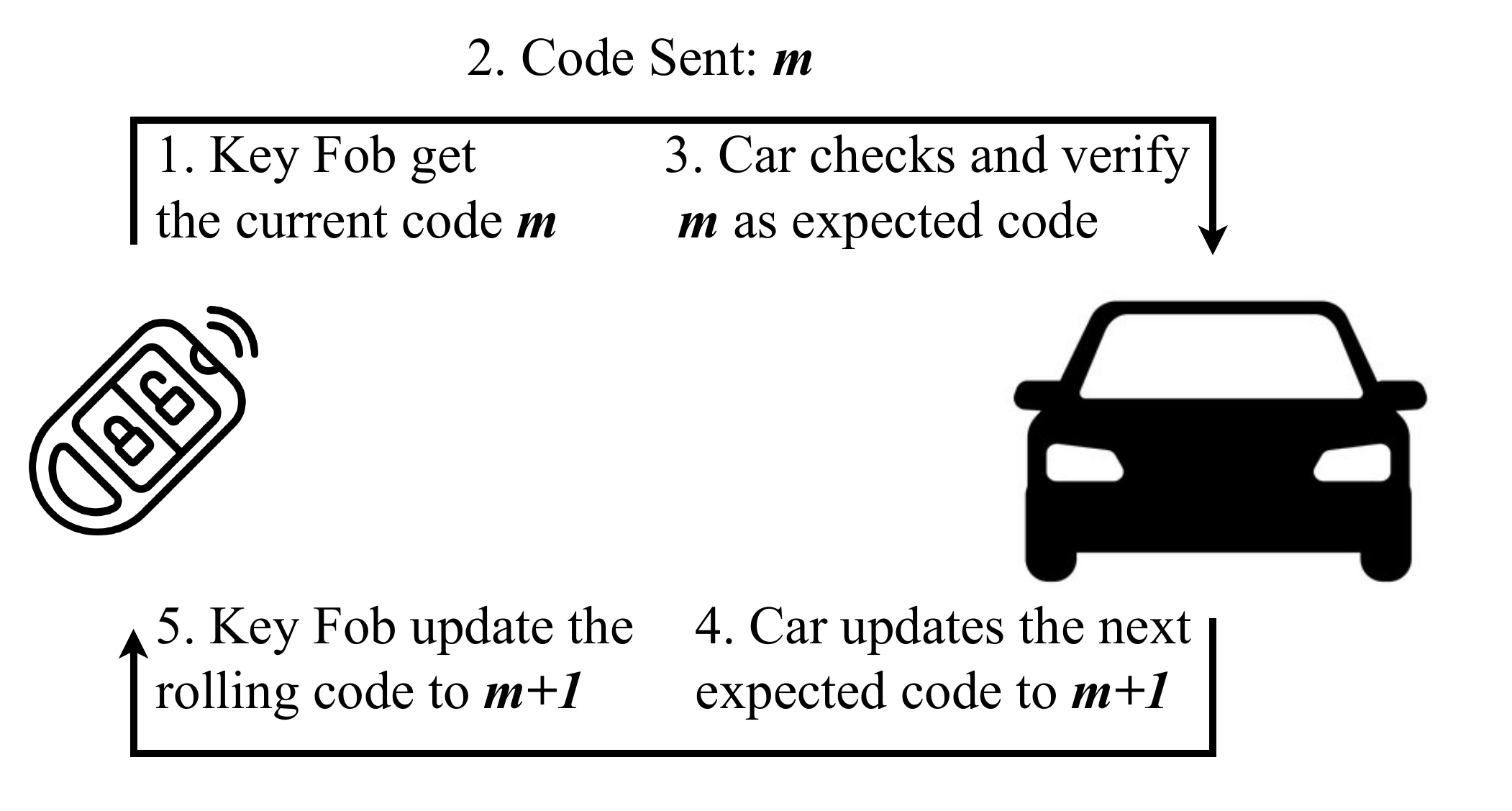}
\end{center}
\caption{\label{fig:rolling_codes} Basic example of Rolling Code usage, where the two parties updated the expected code at each iteration.}
\end{figure}

In the beginning, the key fob used static codes during the communication. 
% As the reader can imagine, t
These static codes were prone to replay attacks due to the identical message transmitted with a press of the same button~\cite{alradaby_fixedcode}.
Thieves could easily steal cars by listening, recording a single button press, and retransmitting the signal.
To avoid exposure to these attacks, manufacturers proposed using rolling codes.
Rolling codes use a synchronized counter in the transmitter and receiver to generate one-time unique signals~\cite{marneweck_introduction_nodate}. 
The transmitter includes the unique code in the message, and the receiver's side compares it against its local counter.
If the values match, the receiver considers the signal valid, and both the transmitter and receiver increment their counter value.
The car has a tolerance range to accept and re-synchronize the counter to foresee accidental button presses.
Figure~\ref{fig:rolling_codes} represents a general and straightforward rolling code functionality. 
Nowadays, some rolling codes replace the incremental function with a more secure \ac{prng} with a shared secret seed between the receiver and transmitter~\cite{marneweck_introduction_nodate}. 
These systems are thus more secure than static code, as they rely on cryptographic functions and usually closed-source schemes built internally by the manufacturers.
However, they are prone to vulnerabilities, and, as we discuss later, attackers exploit them to gain access and take control of the vehicles.

%-------------------------------------------------------------------------------
\section{RKE Implementations}\label{sec:implementation}
%-------------------------------------------------------------------------------
In this section, we describe the evolution and implementations of the technologies for the \ac{rke} and \ac{pkes} in more detail. 
We divide them into legacy technologies, based on rolling codes, and newer ones, using different physical channels and stacks, such as Bluetooth and web applications. 
Table~\ref{tab:rke_types} summarizes the systems, the technology they provide, and in which component we can find them.
The complete and detailed description of each system is out of the scope of this paper, and we refer to the bibliography for additional information.
Yet, we provide the relevant specifications to understand the weaknesses exposing the \ac{rke} systems to attacks.

\begin{table*}[h!]
    \centering
  \caption{Vehicle remote entry technologies and evolution. The first five rows represent rolling code schemes, while after that, we saw the implementation of other protocol stacks, such as \ac{nfc} and web services for vehicle connection to the internet. The acronyms are RKE - Remote Keyless Entry, PKES - Passive Keyless Entry and System, IMM - Immobilizer. The symbol \CIRCLE\, indicates in which component the system is mounted.}
  \label{tab:rke_types}
  \begin{tabular}{|p{4cm}|p{8.5cm}|c|c|c|}
    \hline
    \multirow{2}{*}{\textbf{System}} & \multirow{2}{*}{\textbf{Technology}} & \multicolumn{3}{c|}{\textbf{Component}} \\
    \cline{3-5}
    & & \ac{rke} & \ac{pkes} & IMM\\
    \hline
     \acrshort{dst}40  & 40-bit proprietary challenge-response protocol working over \ac{rfid} from Texas Instruments & & & \CIRCLE \\
     \hline
     KeeLoq  & Proprietary Non-Linear Feedback Register Rolling-Code application& \CIRCLE & & \CIRCLE \\
     \hline
     Hitag2  & Proprietary stream cipher working on a 48-bit key from NPX semiconductor  & & & \CIRCLE  \\
     \hline
     Megamos   & 96-bit secret on a proprietary stream cipher & & & \CIRCLE \\
     \hline
     \acrshort{dst}80  & Texas Instruments 80-bit DST improvement with signature & & & \CIRCLE \\
     \hline
     \acrshort{nfc}-based  & Set of protocols for communication in short distance to open the vehicle & & \CIRCLE & \CIRCLE \\
     \hline
     \acrshort{ble}-based  & Wireless Personal Area Network to connect to the vehicle & & \CIRCLE & \\
     \hline
     \acrshort{uwb}-based  & Radio technology for low energy signals and higher data rates for vehicle-\ac{rke} communication & & \CIRCLE & \CIRCLE \\
     \hline
     \acrshort{api}-based  & Handlers for different actions requested by applications to communicate with the vehicle through the internet connection & \CIRCLE & \CIRCLE & \\
     \hline
     Third-Party Applications  & Applications to provide better control or services with car manufacturers & \CIRCLE & \CIRCLE & \\
    \hline
    \end{tabular}
\end{table*}

\subsection{Legacy}
The majority of the older techniques for \ac{rke} systems are based on cryptographically insecure schemes.
Researchers and attackers identified their weaknesses over the years, significantly improving their security.
Unfortunately, some are still present in key fobs nowadays, and attackers can leverage these systems for illicit car entry.

\paragraph{\textbf{DST40}.} A \ac{dst} is an \ac{rfid} device supplying cryptographic functionalities for authentication.
Texas Instruments designed it to provide security in vehicles via immobilizers, thus requiring authentication before ignition~\cite{dst_bono}.
It is a passive device; therefore, it receives its power via electromagnetic induction. 
This device's peculiarity is that it uses a 40-bit cryptographic key, programmable via \ac{rf} commands.
The \ac{dst} emits an identifier of 24 bits and authenticates with a challenge-response protocol, with a challenge of 40 bits and a truncated 24-bit response.
The security of this device depends entirely on the secrecy of the key.
Due to the low number of bits used, the vulnerabilities found in this system forced companies to adapt the \ac{rke} systems, increasing the number of bits (see \ac{dst}80) or with new ciphers.

\paragraph{\textbf{KeeLoq}.} KeeLoq is a block cipher used in \ac{rke} systems and immobilizer. It consists of a 32-bit block size and a 64-bit key.
KeeLoq uses two registers: key register and state register. 
It can be viewed as a \ac{nlfsr} that depends on the above two registers.
At each iteration, the registers are shifted to the right, and an XOR function computes the new bit; after 528 cycles, the ciphertext is in the state register~\cite{eisenbarth_keeloq_power_2008}.
To authenticate the transmitter, KeeLoq provides a challenge-response protocol in rolling code applications with two keys: the device key, shared and unique to each pair of transmitter and receiver, and the manufacturer key used to derive the device keys.

\paragraph{\textbf{Hitag2}.} It is a stream cipher used in immobilizers from different car manufacturers.
It is a proprietary cipher, but researchers reversed it in  2007~\cite{wiener2007reverse_hitag2}.
Hitag2 implements a 48-bit key as the structure of the internal length of the scheme~\cite{stembera_hitag2_2011}.
The chips contain eight pages of 32 bits for 256 bits in total. 
It can work in distinct modes depending on the objective: read-only, writable (\textit{Password mode}), and \textit{Crypto mode} for additional security.
The \textit{Crypto mode} is the one implemented on vehicles' key fobs.
In this mode, the transponders share a secret 48-bit key, with a unique \ac{iv} generated for each transmission to avoid replay attacks.
Together with the serial number of the tag transponder, the \ac{iv} and the key initialize the cipher that produces a keystream, with the first 32 bits used for authentication and the other 16 bits for encryption.

\paragraph{\textbf{Megamos}.} Car manufacturers used Megamos Crypto in the immobilizer.
Introduced by EM Microelectronic-Marin SA, it consists of a 96-bit secret on a proprietary cipher, with a 32-bit code needed to write in memory.
The authentication proceeds with the car sending a random nonce to the fob with a car authenticator (identifier), and the key fob responds with its authenticator~\cite{verdult_cryptanalysis_2015_megamos}.
The authors in~\cite{verdult_megamos_usenix_2013} reversed the cryptographic scheme, identifying a stream cipher with five main components working as a \ac{prng}.

\paragraph{\textbf{DST80}.} After the problems found in \ac{dst}40, in 2012 Texas Instruments presented the newer \ac{dst}80~\cite{dst80_texas}.
It uses an 80-bit authentication key stored in two halves and can support fast or mutual authentication~\cite{wouters_dst80}.
The authentication protocol follows one of its predecessors with a challenge-response scheme but also includes a block check and a signature.
\\

In legacy systems, the implementation of weak cryptographic measures dictated the development of new cryptographic tools to overcome the previous problems. 
The \ac{rke} systems, independently from the ciphers used, suffer from the physical layer attacks such as relay, jamming, and replay, as we discuss in later sections. 
The reason is that these attacks work directly at a signal level without breaking or interfering with the encryption mechanisms.
The impossibility of overcoming these vulnerabilities led to adopting new technologies that mitigate the problem, such as Bluetooth and \ac{uwb} communication.

\subsection{New Technologies}
With the advent of new technologies, car manufacturers found opportunities to introduce them in their products.
Different \ac{rke} nowadays implement the use of \ac{nfc} with cards or Bluetooth through a smartphone application.
As with any other technology, the comfort they bring comes with the price of decreased security if not properly and carefully implemented.
Nonetheless, these systems promise to reduce the risk of attacks and defend against physical layer attacks that, as previously mentioned, are harder to mitigate due to their nature (relaying, replaying, and jamming).

\paragraph{\textbf{NFC}.} 
It is a set of protocols to enable communication between two devices in a short-range scenario.
Different standards extend the \ac{rfid}~\cite{nfc_specifications}.
Tesla Motors introduced this car access method on the Tesla Model 3~\cite{tesla_nfc}.
To open the car, the driver needs to tap the \ac{nfc} card against the card reader on the door pillar on the driver's side. 
After that, the driver has two minutes to start the car, or it must be re-authenticated by placing the card on the internal reader.

\paragraph{\textbf{\ac{ble}}.}\ac{ble} is a well-known wireless personal area network radio interface technology. 
It is designed for energy-constrained devices and establishes a client-server (phone/car) connection.
Tesla introduced it as a vehicle key via its application~\cite{tesla_nfc}.

\paragraph{\textbf{\acrfull{uwb}}.}It is a radio technology for short-range communications using a wide bandwidth to send signals with very low energy. 
It uses different channels splitting the bandwidth into smaller chunks to favor higher data rates~\cite{bmw_uwb}.
In 2021, BMW announced and later implemented \ac{uwb} in its digital keys with increased security and comfort~\cite{bmw_introduction_uwb}. 
In their statement, they promise security against the infamous relay attack, which we discuss in Section~\ref{sec:attacks}.
\\

As we describe in the next section, the use of these technologies does not fully mitigate the possibility of attacks to \ac{rke} and \ac{pkes} systems.
The introduction of web and smartphone applications can not only increase the usability and comfort of the users but also help in defending against theft and attacks by sending messages through these channels.

\subsection{Advent of the Web}
In recent years, the automotive sector has seen the opening to the web as a tool for collecting vehicle data and diagnostics. 
Nowadays, the web is also used to provide new smart features to cars and their owners. 
Opening a car and starting the engine or opening the heating is possible through the touch of a button on a smartphone.
This introduces a new attack surface that largely increases the possibility of errors for developers and the exploitation by attackers, as in the everyday web security field~\cite{web_security}.

\paragraph{\textbf{\ac{api}}.}The use of this web application system requires the implementation of \ac{api}s to handle the different actions~\cite{car_api}. 
These are not limited to opening or closing the doors but are also used by manufacturers to query the car for diagnostic data. 
Different car manufacturers have closed \ac{api}s, but researchers are working on reversing them as it happens with Tesla~\cite{tesla_api} or Kia~\cite{kia_api}.

\paragraph{\textbf{Third Party Apps}.}To provide better control of cars, different manufacturers allow the integration of third-party applications, also integrating additional \ac{api}s.
The drivers are turning towards these providers because they offer better features at a lower price, sometimes also coming with transparency on the data usage~\cite{why_third_party}.
Also, in this case, introducing third parties in the system increases the attack surface available for malicious users~\cite{tesla_third_party, third_party_apis}.
\\

Even if these systems can help prevent attacks and increase the complexity of possible exploitations, the web introduces all the risks associated with it, such as the normal web exploitation vectors, as we describe in the next section.

%-------------------------------------------------------------------------------
\section{How They Steal Your Car: Attack Strategies}\label{sec:attacks}
%-------------------------------------------------------------------------------
In this section, we describe the macro-area of attacks against \ac{rke} systems and the exploits implemented in real-world scenarios during the years, showing the similarity between the first attacks and the upcoming ones in these last few years. 
Table~\ref{tab:attacks} contains all the attacks divided by type and in chronological order, indicating the target of the exploit.
As we can see, cryptographic attacks highly focus on retrieving the secret key from the devices.
We left out lock picking from the attacks due to the physical and not \ac{rke} related nature.

The different areas consist of the kind of attack and means to carry it.
Specifically, we have cryptographic attacks and different wireless signal methods to intercept and break the \ac{rke} systems.
Some attacks combine the different classes to exploit weaknesses in a cipher, retrieve the key craft authenticated packets, or perform other attacks subsequently. 
Section~\ref{subsec:crypto-attack} describes the various cryptographic attacks presented in the literature. 
In Section~\ref{subsec:relay-attack} and ~\ref{subsec:rolljam-attack}, we present the attacks based on radio technologies such as jamming and replay, while in Section~\ref{subsec:newtech-attack}, we show how new technologies employed in these systems still present weaknesses and introduce an exploitable point leveraged by attackers.
Finally, in Section~\ref{subsec:web-attack}, we describe the weaknesses of introducing web technologies into the automotive industry related to the remote keyless entry.

\begin{table*}
  \centering
  \caption{The attacks targeting \ac{rke} and \ac{pkes} systems grouped by the attack class. The acronyms for the attack targets and goals are as follows: RKE: Remote Keyless Entry, PKES: Passive Keyless Entry and Start, IMM: Immobilizer, AC = Action on Car (open the doors or start the car), SK = Secret Key retrieval, CK = Clone the Key fob. The \CIRCLE symbol means an attack against a target and following a specific goal. In contrast, the \LEFTcircle symbol indicates if it is related to that domain due to a side effect (e.g., the attack targets the RKE system but can also be used to start the car through the immobilizer bypass).}
  \label{tab:attacks}
  \begin{tabular}{|p{0.4cm}p{2.8cm}|p{6cm}|c|c|c|c|c|c|}
    \hline
    \multicolumn{2}{|c|}{\multirow{2}{*}{\textbf{Attack Class}}} & \multicolumn{1}{c|}{\multirow{2}{*}{\textbf{Attack}}} & \multicolumn{3}{c|}{\textbf{Target}} & \multicolumn{3}{c|}{\textbf{Goal}} \\
    \cline{4-9}
     &  &  & \ac{rke} & \ac{pkes} & IMM & AC & SK & CK \\
    \hline
    \multirow{16}{*}{\rotatebox{90}{Cryptographic}}  &  \multirow{5}{*}{Exhaustive Search} & Analysis of DST40 \cite{dst_bono} & \LEFTcircle &  & \CIRCLE & & \CIRCLE & \LEFTcircle \\
    &  & Hitag2 hardware optmization\cite{stembera_hitag2_2011}  &  & & \CIRCLE &  & \CIRCLE &  \\
    &  & Megamos Crypto \cite{verdult_megamos_usenix_2013,verdult_cryptanalysis_2015_megamos}  &  &  & \CIRCLE &  & \CIRCLE &  \\
    &  & Hitag2 optimized attack\cite{benadjila_hitag2_2017}  &  &  & \CIRCLE &  &  \CIRCLE & \\
    &  & Analysis of DST80\cite{wouters_dst80_2019}  &  & \CIRCLE &  &  & \CIRCLE & \CIRCLE \\
    \cline{2-9}
    
    & \multirow{2}{*}{GD} & KeeLoq Cryptanalysis\cite{Bogdanov2007KeeLoq} & \CIRCLE &  & \CIRCLE &  & \CIRCLE &   \\
    &  & Hitag2 optimized GD\cite{vergesten_gd_hitag2_2018}  & &  & \CIRCLE &  & \CIRCLE &   \\
    \cline{2-9}
        
    & \multirow{2}{*}{Slide Attack} & KeeLoq Algebraic and Slide attack\cite{courtois_slide_keeloq_2008} & \LEFTcircle &  & \CIRCLE &  & \CIRCLE &  \\
    &  & Practical attack on KeeLoq\cite{indesteege_slide_hitag2_2008}  & \CIRCLE &  & \CIRCLE  &  & \CIRCLE &  \\
    \cline{2-9}
    
    & \multirow{2}{*}{Algebraic Attack} & KeeLoq Algebraic and Slide attack\cite{courtois_slide_keeloq_2008} &  & \CIRCLE &  &  & \CIRCLE &  \\
    &  & Hitag2 practical algebraic attack\cite{courtois_algebraic_hitag2_2009}  &  &  & \CIRCLE &  & \CIRCLE &  \\
    \cline{2-9}
    
    & \multirow{2}{*}{Correlation Attack } & Hitag2 dependencies between sessions \cite{verdult_gone_in_360_hitag2} & & & \CIRCLE &  & \CIRCLE &  \\
    &  & Insecurity of Hitag2 RKE systems\cite{garcia_hitag2_correlation_2016}  & \CIRCLE &  & \LEFTcircle & \LEFTcircle & \CIRCLE & \CIRCLE  \\
    \cline{2-9}
    
    & \multirow{2}{*}{Power Analysis} & KeeLoq Code Hopping\cite{eisenbarth_keeloq_power_2008} & \CIRCLE &  & \CIRCLE & & \LEFTcircle & \CIRCLE \\
    &  & Extarcting KeeLoq keys\cite{kasper_keeloq_power_2009}  & \CIRCLE & & \CIRCLE &  & \CIRCLE & \\
    &  & Dismantling the DST80 immbolbilizer \cite{wouters_dst80_power_2020}  &  & \CIRCLE &  &  & \CIRCLE &   \\

    \hline
    \multicolumn{2}{|c|}{Relay Attack} & Relay on PKES systems\cite{francillon_relay_2011} & \CIRCLE & \CIRCLE & \CIRCLE & \CIRCLE &  &  \\

    \hline
    \multicolumn{2}{|c|}{\multirow{5}{*}{Jamming and Replay Attack}} & RollJam\cite{rolljam} & \CIRCLE & \CIRCLE & \CIRCLE & \CIRCLE &  &  \\
     & & RollBack\cite{levente_rollback_blackhat_2022, levente_rollback_2024} & \CIRCLE & \CIRCLE & \CIRCLE & \CIRCLE &  &   \\
     & & RollingPwn\cite{rollingpwn_2022}  & \CIRCLE & \CIRCLE &  & \CIRCLE &  &   \\
     & & RollJam with known noise source\cite{depp_rolljam_enhance_2023}  & \CIRCLE & \CIRCLE & \CIRCLE & \CIRCLE &  &   \\
     & & RollJam revisited\cite{satish_rolljam_2024}  & \CIRCLE & \CIRCLE & \CIRCLE & \CIRCLE &  &   \\
     
    \hline
    \multicolumn{2}{|c|}{\multirow{5}{*}{Bluetooth and \ac{nfc}}} & TESLA Model 3 NFC Relay\cite{nfc_relay_2020} &  & \CIRCLE &  & \CIRCLE &  &   \\
    & & Tesla Model X PKES compromising\cite{wouters_tesla_pkes_2021} & & \CIRCLE &  &  &  & \CIRCLE  \\
    & & Ghost Peak: UWB distance reduction\cite{leu_uwb_2022} & \CIRCLE & \CIRCLE & \CIRCLE & \CIRCLE &  &  \\
    & & BLE Phone-as-a-key Relay\cite{ble_tesla_relay_2022} &  & \CIRCLE &  & \CIRCLE &  &   \\
    & & Tesla Model 3 RKE compormising\cite{xie_tesla_ble_2023} & \CIRCLE & \CIRCLE &  &  &  & \CIRCLE  \\
    
    \hline
    \multicolumn{2}{|c|}{\multirow{6}{*}{Web Services}} & Remote Started admin API\cite{jmaxxz_api_2019} &  &  & \CIRCLE & \CIRCLE &  &  \\
    & & Exploitation of Honda Connected App\cite{honda_api_2002} & \CIRCLE &  & & \CIRCLE &  &  \\
    & & SiriusMX API\cite{siriusmx_api_2022} & \CIRCLE & & & \CIRCLE &  &  \\
    & & TeslaMate API\cite{api_teslamate} & \CIRCLE &  &  & \CIRCLE &  &  \\
    & & Tesla Logging Web Service\cite{tesla_log_2024} & \CIRCLE &  &  & \CIRCLE &  &  \\
    & & Kia Service API\cite{kia_api_2024} & \CIRCLE &  &  & \CIRCLE &  &  \\
    
    \hline
    \end{tabular}
\end{table*}

\subsection{Cryptanalytic Attacks}\label{subsec:crypto-attack}
This category includes all the attacks that target the cryptographic functions implemented inside \ac{rke}, \ac{pkes}, or immobilizer. 
Generally, these attacks require access and recording of some messages between the car and keyfob and reverse engineering the algorithm, ultimately allowing the recovery of the key and cloning the keyfob.
Additionally, they affect all the companies and brands using the same type of cryptographic tools inside their keyfobs.

\paragraph{\textbf{Exhaustive search}.} 
The most straightforward approach for the attacker is to search through all the possibilities. 
Depending on the search space (e.g., key size), it may require significant time but eventually converges to a solution. 
The search space can also be decreased by exploiting weak ciphers and keys.
This method allows the attacker to find the key for the \ac{rke} system. 
The first attack in this category is the cracking of \ac{dst}40 by~\cite{dst_bono}.
The authors reversed the protocol and needed only two challenge-response pairs to recover the keys. 
Over the years, different exhaustive search attacks allowed hackers to steal the \ac{rke} and \ac{pkes} keys.

After~\cite{dst_bono}, researchers broke the (in-)famous Hitag2 in 2011~\cite{stembera_hitag2_2011} and~\cite{benadjila_hitag2_2017}. 
The first attack details a new method for breaking the stream cipher with only two sniffed messages and significantly less time than the available algebraic attack: 2 hours instead of 45, thanks to the hardware implementation on COPACOBANA platform.
In~\cite{benadjila_hitag2_2017}, the authors performed a black-box analysis. They identified a weakness in the initialization vector that allows an attacker to find the key in minutes with access to a cluster to perform an optimized brute force search.

Even the other ciphers, such as Megamos and \ac{dst}80 were not secure, as Verdult et al.~\cite{verdult_megamos_usenix_2013} and Wouters et al.~\cite{wouters_dst80_2019} showed in their works.
The first work presents the inner functionalities of Megamos Crypto, showing its weaknesses and how to exploit them to recover the 96-bit key with a computational complexity of $2^{56}$ cipher ticks (or $2^{49}$ encryptions).
They target various car makers and models, finding they used weak keys (only ten over 96 bits were ones) that allowed for a partial key-update attack with additional optimization.
In 2019, Wouters et al. reversed the \ac{dst}80 cipher used in the Tesla Model S vehicles \ac{pkes}, revealing only 40 bits system, lack of mutual authentication, and other memory vulnerabilities~\cite{wouters_dst80_2019}.
These vulnerabilities allowed the attackers to clone the key fob quickly from a challenge-response round with a \ac{mitm} attack, employing a time-memory trade-off technique to reduce the computation time to recover the key. 
    
\paragraph{\textbf{\acrfull{gd}}.}
\ac{gd} is a technique used in cryptanalysis to recover unknown variables in a system.
It consists of guessing a subset of the variables in order to deduce the remaining values and their relationship~\cite{vergesten_gd_hitag2_2018}.
This is particularly useful when the cipher does not use the whole internal state to compute the keystream (stream ciphers). 
The attacker can initially guess only a partial internal cipher state and evaluate the output, drastically reducing the search time compared to the exhaustive methodology.
The first work using this method against \ac{rke} systems attacks the KeeLoq and its self-similar key schedule to guess a portion of the key using the sliding technique to generate pairs of input and output~\cite{Bogdanov2007KeeLoq}.
Ultimately, the author recovered the entire key by deducing linear relationships with the key bits with a complexity of $2^{50.6}$ (with a more involved attack with $2^{37}$ complexity) with the requirement of $2^{32}$ known plaintext-ciphertext pairs. 
Researchers also used the \ac{gd} technique to break the Hitag2 ciphers.
Vergesten et al. improved the \ac{gd} attack on Hitag2 key recovery through different optimizations~\cite{vergesten_gd_hitag2_2018}.
The authors claim an improvement of over 500 times with respect to the previous fast Hitag2 attack ~\cite{benadjila_hitag2_2017}.
These attacks allow the recovery of the keys and, ultimately, from the reverse engineering of a single internal state, clone the keyfob.

\paragraph{\textbf{Slide attack}.}
This cryptanalysis technique aims to break a cipher in multiple rounds identified in identical function representations~\cite{slide_attack}.
A slide attack deals with ciphers with a high number of rounds, rendering complexity-based security ineffective. 
In the literature, two works from 2008 focus on slide attacks against KeeLoq.
The first work, proposed by Courtois et al.~\cite{courtois_slide_keeloq_2008}, combines the algebraic and slide attacks to noticeably simplify the key recovery by guessing 16 bits, corresponding to the last 16 rounds and thus reducing the total number from 528 to 512 with a total complexity of $2^{53}$ KeeLoq encryptions.
The drawback is the requirement of $2^{16}$ known plaintexts.
In the second paper, Indesteege et al.~\cite{indesteege_slide_hitag2_2008} reduced the complexity to $2^{44.5}$ KeeLoq encryptions with $2^{16}$ chosen plaintexts (same requirement of the previous attack) thanks to a new meet-in-the-middle approach.
They fully implemented the attack, which recovered the master key in less than five minutes and replicated the full device.

\paragraph{\textbf{Algebraic attack}.}
It expresses the cipher as algebraic equations, replacing the known data in the system and trying to search for the key. 
The technique is highly successful against ciphers using linear operations, while systems combining non-linear functions make this method harder to utilize.
In \cite{courtois_slide_keeloq_2008} and \cite{courtois_algebraic_hitag2_2009}, Courtois et al. proposed two algebraic attacks first against KeeLoq, described before in the Slide attack paragraph, and against Hitag2.
In both works, they leveraged SAT solvers to represent and solve the system of multivariate equations, guessing some variables and examining the consequences.
In \cite{courtois_algebraic_hitag2_2009}, they focused on the challenge-response Crypto mode of Hitag2, with a Known IV attack to make it practical.
It requires data from 4 transactions, guessing 14 bits of the key, and combining the equations for the four known IVs.
It requires around two days for the full 48-bit key, but it is fully automated and can work against Hitag2 with slight modifications.
    
\paragraph{\textbf{Correlation attack}.}
A correlation attack is used against stream ciphers that combine different \acp{lfsr}, using their output as boolean functions. 
This attack exploits the correlation arising from statistical weaknesses.
The two attacks in this category represent some of the main works against \ac{rke} systems exploiting cryptographic weaknesses in Hitag2.

"Gone in 360 Seconds", by Verdult et al.~\cite{verdult_gone_in_360_hitag2}, uses the lack of \ac{prng} and redundancy as an arbitrary length keystream oracle.
The authors also observed a one-bit leak in the secret key for every four authentications and 16-bit persistent information in the secret key over multiple sessions.
They presented two attacks: the first one uses the first vulnerability and uses a keystream shifting attack, while the second one is against the \ac{lfsr} and is more general. 
Their evaluation of 20 different cars found that the \ac{lfsr} seed was based on time, introducing weak and predictable (or even default) passwords and low entropy keys.
With these vulnerabilities, they were able to use correlation to predict possible keys and perform a dictionary attack due to the low entropy of the keys.

The second main work by Garcia et al.~\cite{garcia_hitag2_correlation_2016} analyzed the Volkswagen group, finding that they used only a few global master keys.
In this way, they could clone remote controllers by eavesdropping only one signal.
Against Hitag2, they described a novel correlation attack, called \textit{fast-correlation attack}, recovering the key with only 4 to 8 signals in one minute of computation.
To do so, they make guesses on the counter window, which initializes the cipher with a 28-bit counter.
They found that only 10 bits were used over the air by reverse engineering; therefore, the attacker needs to guess the remaining 18 bits.
After that, the attacker clones the key fob and can access the vehicle for further exploitation.

\paragraph{\textbf{Power Analysis}.}
Power analysis consists of a side-channel attack aiming to extract the secret key by analyzing the power consumption of the hardware during cryptographic operations.
Power analysis against \ac{rke} systems focused specifically on KeeLoq in 2008 and 2009~\cite{eisenbarth_keeloq_power_2008, kasper_keeloq_power_2009}, with a recent work against \ac{dst}80~\cite{wouters_dst80_power_2020}.
Eisenbarht et al. \cite{eisenbarth_keeloq_power_2008} demonstrated that ten power traces are enough to clone a remote controller and extract the manufacturer key with differential power analysis.
Once the attacker knows the manufacturer key and key derivation algorithm, it can also perform power analysis on two hopping code messages eavesdropped from the remote.
This is possible because the KeeLoq remotes analyzed do not use any seed for key derivation.
In~\cite{kasper_keeloq_power_2009}, the authors use simple power analysis to extract the key, drastically reducing the time required (seconds) for the operation with respect to the previous attack.
In this case, only one measurement is enough to extract the 64-bit master key.
Concerning power analysis in this environment, there is a more than ten-year gap until Wouters et al. published an attack against \ac{dst}80 in 2020\cite{wouters_dst80_power_2020}.
They exposed significant flaws in key diversification schemes used by major car manufacturers like Toyota, Kia, and Hyundai, revealing extremely low entropy in the generated keys.
In this paper, the authors used voltage glitching to extract the firmware and side-channel attacks to recover the cryptographic key.
The same authors in ~\cite{wouters_dst80_2019}, unveiled the incorrect use of \ac{dst}80 as identical to the \ac{dst}40, thus inheriting the older vulnerabilities and weaknesses.

As we described, some cryptographic attacks combine multiple techniques to reach the goal of extracting the secret keys and cloning the key fob.
Nowadays, these attacks are less applicable due to the change in the \ac{rke} and \ac{pkes} systems.
Nonetheless, security vulnerabilities are always a threat, especially in cryptographic closed-source functions that lack proper testing.

\subsection{Relay Attacks}\label{subsec:relay-attack}
\begin{figure}
\begin{center}
\includegraphics[width=\columnwidth]{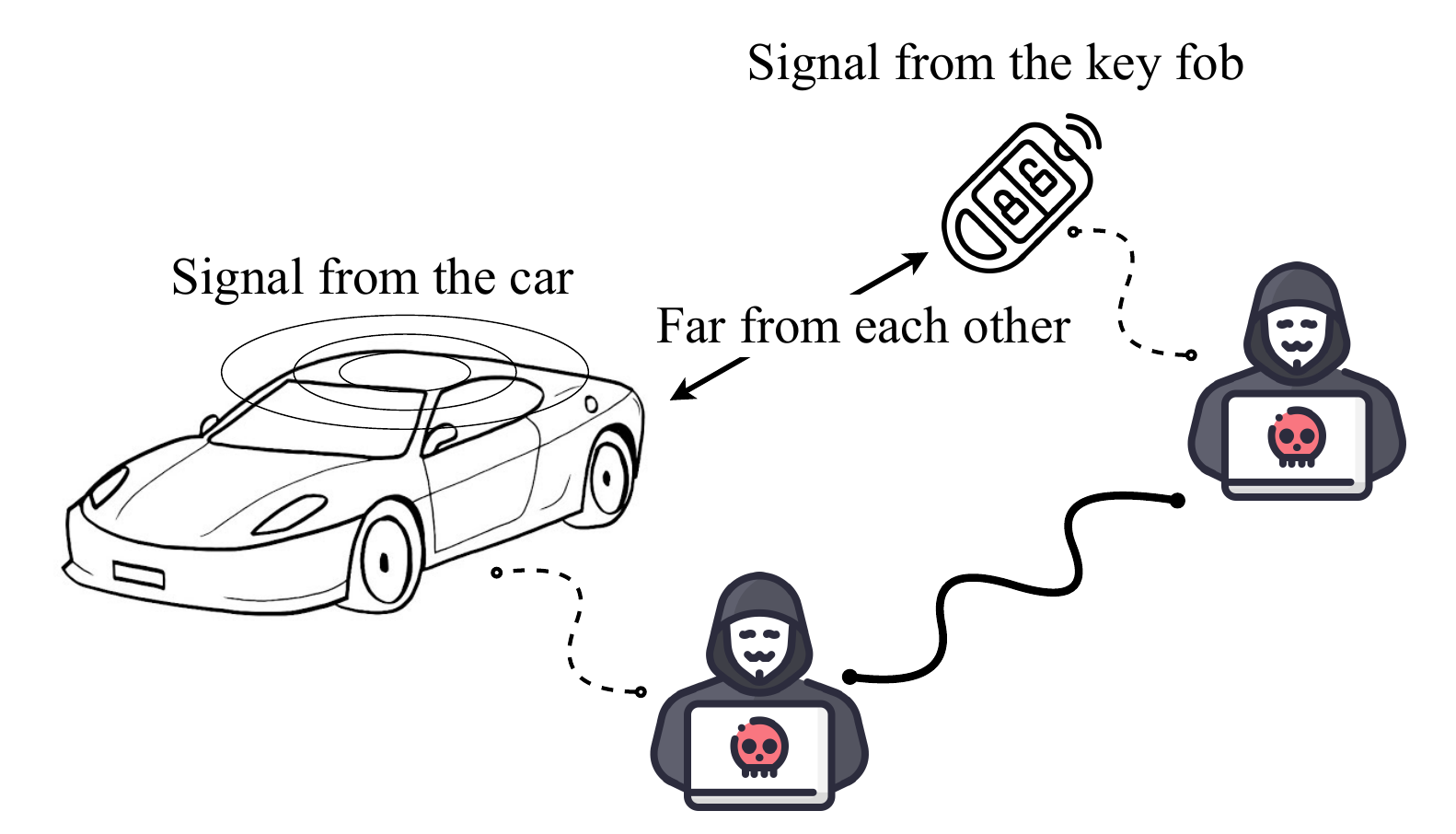}
\end{center}
\caption{\label{fig:relay} Relay attack representation where the two bad actors intercept the signals from the vehicle and the key fob, tricking them into thinking they are in close range.}
\end{figure}

A relay attack is a \ac{mitm} combined with a replay attack. 
Specifically in \ac{rke} scenario, the attacker manipulates the communication between the transceiver on board the vehicle and the key fob, intercepting and forwarding the message in an intermediary channel to cover the distance separating the two devices~\cite{relay_explanation}.
In this way, the car transceiver thinks to be in the area of the key fob and to communicate with it.
The same happens in the opposite direction, with the key fob believing being close to the car range.
In Figure~\ref{fig:relay}, we show the steps for the attackers (usually two partners) to relay the signal and steal a car.
This attack targets all the legacy systems independently of the model and car manufacturer.

Francillon et al., back in 2011, proposed the relay attack against vehicles and \ac{pkes} systems~\cite{francillon_relay_2011}.
They built the attack in the experimental scenario with 10 cars measuring the distance reachable to open the vehicle, relaying both the \ac{lf} and \ac{uhf} signals of the \ac{pkes}, also with amplification.
They found that without amplification, they could reach up to 2 meters from the car, while with an amplified signal, this range extended up to 8 meters. 
On the opposite, the distance from the key fob, due to the use of higher frequencies, could be up to 60 meters.
The authors note that the main reason for the success of this attack is the verification of communication with the correct key, but not that the correct key is in proximity.
The communication condition is (wrongly) assumed to be verified if the two devices communicate, thus deeming only key verification necessary.
As this attack shows, this wrong assumption makes the exploitation possible, and the \ac{pkes} systems are still vulnerable due to the same weakness~\cite{greenberg_teslas_nodate, ungoed_thomas_gone_2024}. 

\subsection{Jamming and Replay}\label{subsec:rolljam-attack}
\begin{figure}
\begin{center}
\includegraphics[width=\columnwidth]{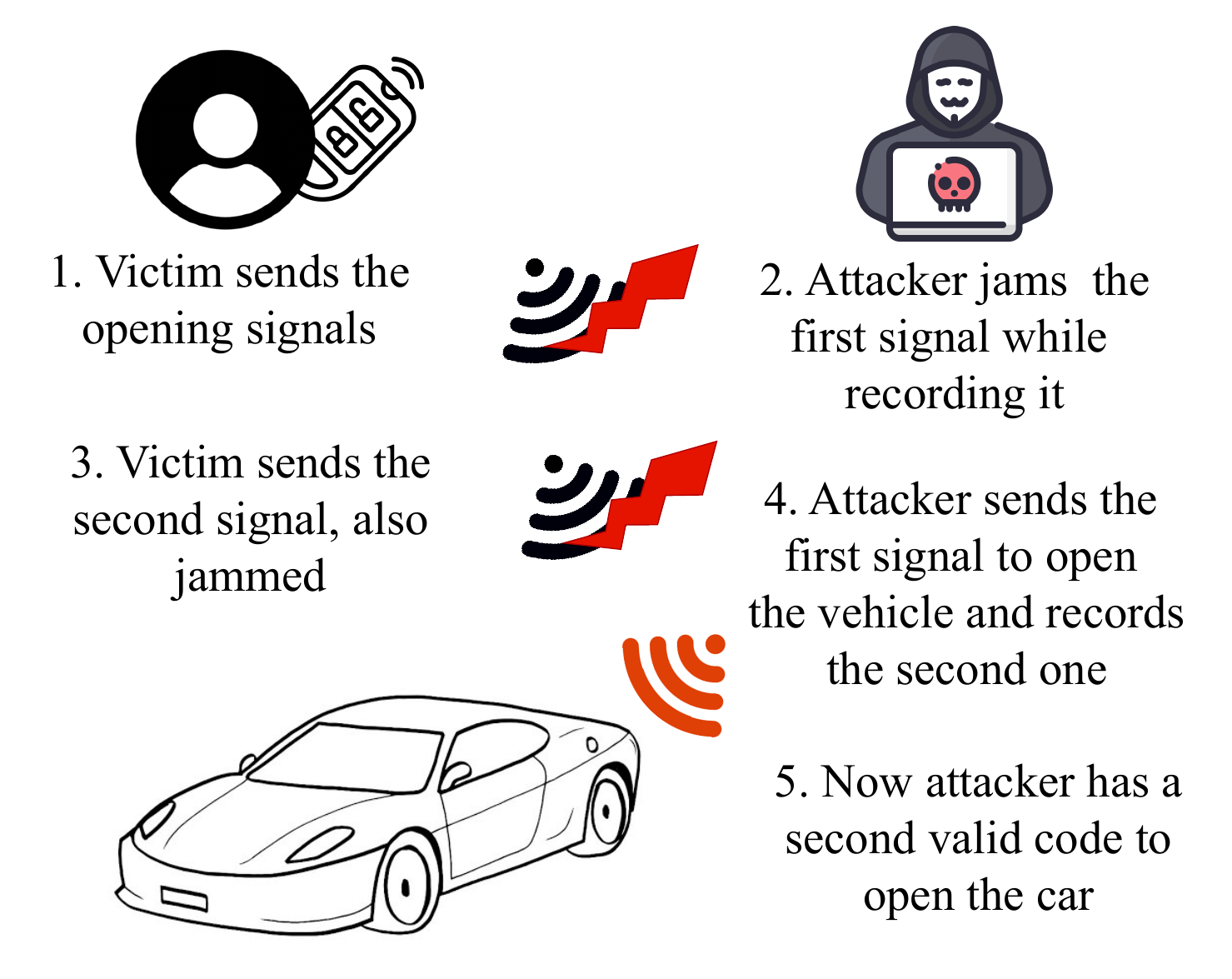}
\end{center}
\caption{\label{fig:rolljam} Rolljam attack steps representation, in which an attacker jams and sniffs the two signals from the victim while sending the first capture during the second jamming. 
% In this way, the car appears to open with the second button press, and the attacker gains a valid code for a future moment.
}
\end{figure}

The Jamming and Replay attack, also called RollJam, was first introduced in 2015 by the security researcher Samy Kamkar~\cite{rolljam}.
The attack requires very low-budget hardware as \ac{sdr} and targets the \ac{rke} and ignition of cars using rolling codes.
It consists of recording and blocking the radio signal from the key fob when the driver tries to unlock the doors.
The driver will try again to unlock the vehicle. 
Still, the attacker will execute the jamming and recording against the second signal while replaying the first recorder message to let the driver think the unlock worked the second time.
Now, the attacker has a second, valid code to unlock the car, and the driver will lock it the next time.
In Figure~\ref{fig:rolljam}, we represent the attack steps.
Starting from this attack, different researchers improved the procedure or proposed alternative approaches.
Also, many people tried it against various vehicle brands and models, confirming and proving the exploitability of this attack in the wild~\cite{ayyappan_rolljam_2024, ibrahim_rolljam_2019, urquhart_rolljam_2019, satish_rolljam_2024, rolljam_demo_2024}.
Especially these last works present simple solutions to perform the RollJam attack and analyze and test \ac{rke} systems automatically~\cite{satish_rolljam_2024}.

The first work targeting the vehicle opening mechanism is RollingPwn~\cite{rollingpwn_2022}.
The ten most popular Honda vehicles from 2012 to 2022 are affected by this vulnerability of the rolling code mechanism.
The Honda's weakness consists of accepting codes from a rolling code from the previous cycle after resynchronizing the counter.
The authors published the attack on a webpage after Honda Motors stated that this replay method would not work against their key fobs implementing rolling codes. 

The second attack using the Jamming and Replay paradigm is RollBack by Levente et al.~\cite{levente_rollback_blackhat_2022, levente_rollback_2024}. 
They presented a setup similar to RollJam, but they only recorded the second unlock signal instead of jamming it.
After that, the owner can use the car as many times as it wants.
The attacker can replay the two consecutive packets and unlock the car thanks to a resynchronization to a previous state with the first signal.
The authors confirmed the vulnerability of Kia and Mazda vehicles.

The latest work improves the RollJam attack using a known noise source, rebutting the requirement of specific knowledge of the attack surface and \ac{sdr} parameters for the original attack~\cite{depp_rolljam_enhance_2023}.
Using a known noise signal for jamming introduces significant improvements in the attack success rate, in contrast to the additive white Gaussian noise used in the original work.

\subsection{Attacks Against New Technologies}\label{subsec:newtech-attack}
In recent years, multiple car brands have redesigned the \ac{rke} and \ac{pkes} systems to new technologies such as Bluetooth, \ac{nfc}, and \ac{uwb}.
Utilizing these communication technologies made stealing a vehicle harder but not impossible.
In fact, various researchers found weaknesses inherited by the protocol stack or a broken implementation of the communication mechanism.

Starting in 2020, the hacker named Kevin2600 and its Chaos-Security-Lab presented a relay attack against \ac{nfc} opening used by Tesla~\cite{nfc_relay_2020}.
They analyze the internals of the key fob and the \ac{nfc} Tag (in this case, a Java smart card). 
They use an Android application called NFC-Gate to perform a \ac{mitm} attack, reversing the NFC communication and actuating a relay attack against the Tesla with a relay transmission over WiFi. 
They reported the findings to Tesla, who answered without much attention and suggested using the additional security called Pin2Drive (inserting a pin inside the vehicle to start the car). 
They also checked for the pin, which had only four digits, without any brute-force protection. 
Also, in this case, they could still access the car. 

After that, in 2022, researchers in NCC Group found a relay attack on the \ac{ble} link layer~\cite{ble_tesla_relay_2022}.  
It was possible to add latency within the normal GATT response timing variation to make relaying of the encrypted link layer possible.
In addition, this attack could bypass the relay mitigations and bounds accepted by the Tesla Model 3 \ac{pkes}.
Always against Tesla, a year later, Xie et al. reverse-engineered and exploited the Phone Key feature through a \ac{mitm} attack on the Bluetooth channel~\cite{xie_tesla_ble_2023, temsla}.
Additionally, they found a weakness in the Key Card pairing protocol, allowing them to connect a programmable Java card.
Combining the several flaws explained in their paper, the authors could trick the vehicle into believing that the phone used by the attacker was authenticated and registered, thus allowing the attacker to enter and start the car without any awareness by the actual owner.

A work not strictly related to the channel communications presented before, even if passing over \ac{ble}, targets the \ac{pkes} and \ac{bcm} of the Tesla Model X to unlock and start the vehicle~\cite{wouters_tesla_pkes_2021}.
Two vulnerabilities allowed the researchers to compromise the system: the firmware on the key fob did not properly verify the image's authenticity through the \ac{ble} interface, and the pairing protocol never sent the certificates to verify the key fob's authenticity to the car.
Combining the two vulnerabilities, they were able to unlock and start a Tesla in a few minutes, assuming access to the Secure Element in the key fob.
The researchers built a portable device that could carry out the attack, using modified Tesla components to wake up a target key fob, install malicious firmware, and pair a rogue key fob to the vehicle.

Concerning \ac{uwb} technology, car manufacturers started to include it in the new \ac{rke} systems in the last few years, as introduced in Section~\ref{sec:background}.
Since the introduction started around 2021, researchers already found weaknesses even after the promise of \ac{uwb} being secure against relay attacks.
Leu et al. presented the first over-the-air attack on the \ac{uwb} measurement system~\cite{leu_uwb_2022}. 
The authors could reduce the perceived distance in the high-rate physical layer settings, called \ac{hrp}.
In this attack method, the power of the injected packet is selectively varied per packet field to avoid being perceived as a new packet or as jamming. 
The researchers refer to this attack as \textit{selective overshadowing}, in which the attack signal is synchronized with the legitimate signal.
However, the attacker does not need to know a specific packet field for security applications. 
In their experiments, the authors reduced the measured distances from 12 to 0 meters with a success rate of 4\%, sufficient to deceive ranging systems that rely on single \ac{hrp} measurements.
This attack is particularly important for the possible implications in the \ac{rke} scenario.
During the last year, Wired reported the criticalities of \ac{uwb} and their implementations in the automotive industry: Tesla vehicle could still be stolen by the old-fashioned relay attack against Bluetooth due to the absence of using \ac{uwb} for distance checking~\cite{tesla_uwb_wired_2024, tesla_uwb_vicone_2024}.

\subsection{Web Service Exploitation}\label{subsec:web-attack}
The automotive environment is more and more connected to the web realm.
A multitude of services are available for connected cars, including the \ac{rke} remotely managed, for example, through an application on a smartphone.
Different platforms and third-party applications can also interact with vehicles, leaving space for common attacks from the web space.
First, \ac{api}s interacting with the cars can introduce attackers into the system, as shown in the Jmaxxz work presented in 2019~\cite{jmaxxz_api_2019}.
Jmaxxz found a remote starter on the web that uses an application as a remote.
The author found the firmware and where it needed access to the Internet, revealing an \ac{api} key with default user and password and a SQL injection attack to become the administrator.
From there, Jmaxxz could start the car with another SQL injection vulnerability.
A multitude of works targeted the \ac{api} as the remote exploitation of Honda cars in 2022~\cite{honda_api_2002} or the SiriusXM connected vehicle service vulnerability that exposed cars from Honda, Nissan, Infiniti, and Acura~\cite{siriusmx_api_2022, api_2023}.
In both cases, the researchers found vulnerabilities in the applications and \ac{api} that allowed them to bypass account security and execute remote commands, such as unlocking the cars.
In 2023, a security researcher was able to access Tesla cars all over the world thanks to a vulnerability in the \ac{api}s of a third-party application (TeslaMate) used to track the vehicle's movements and perform some actions, including unlocking the doors~\cite{api_teslamate}.
Other vulnerabilities allowed access to the cars through web applications for logging or improper account and request validation~\cite{tesla_log_2024, kia_api_2024}.
Different other vulnerabilities can also give access to private information and actions to malicious users, and different brands are involved, also from major manufacturers such as Rolls Royce and Porsche~\cite{sam_curry_blog}. 
This highlights how the increasing interconnection between automotive and the web is drastically enlarging the attack surface, with targets that can differ from unlocking the vehicle's doors. 

%-------------------------------------------------------------------------------
\section{How Researchers Defend From Thieves}\label{sec:defenses}
% Systematization of Defenses
%-------------------------------------------------------------------------------
This section presents the research and possible defensive techniques available for \ac{rke} and \ac{pkes} schemes. 
The solutions presented differ in the method and the technology adopted, ranging from distance bonding protocol to a newer proposal based on a quantum key distribution protocol to be safe in the post-quantum era.
In Table~\ref{tab:defenses}, we overview the researchers' defense proposals and the attacks they mitigate.
We will not discuss, if not briefly at the end of the section, the immediate-term and mid-term solutions found in the attack papers due to their inadequate usability for the \ac{rke}, such as the battery removal, shielding the key fob (thus a viable option), or kill-switch for the key fob functionalities.
We focus only on the long-term and new paradigm proposals to defend the \ac{rke} system, dividing them based on the mitigation against Replay/RollJam or Relay attacks, except for one work that aims at protecting against power analysis.
Due to the broader scope and general mitigation and defenses for \ac{api}s and web services, we still do not see significant research on web security explicitly related to automotive.

\begin{table*}
  \centering
  \caption{Defenses presented in this work. The acronyms for the target mean C = Crypto, R = Relay, JR = Jamming and Replay, NT = New Technologies, WA = Web Applications. The acronyms for the devices are as before. RKE stands for Remote Keyless Entry, PKSE for Passive Keyless Entry and Start, and IMM for Immobilizer. The \CIRCLE\ symbol indicates which attack the defense wants to mitigate or remediate and the devices it uses. \textit{N/A} indicates the paper does not make values available.}
  \label{tab:defenses}
  \begin{tabular}{|c|c|c|c|c|c|c|c|c|c|c|}
    \hline
    \multirow{2}{*}{\textbf{Defense}} & \multicolumn{2}{|c}{\textbf{Cost (ms)}} & \multicolumn{5}{|c}{\textbf{Target}} & \multicolumn{3}{|c|}{\textbf{Device}}  \\
    \cline{2-11}
    & Key & Car & C & R & JR & NT & WA & RKE & PKSE & IMM \\
    \hline
    RKE power analysis defense\cite{defense_power_2009} & N/A & N/A & \CIRCLE & & & & & \CIRCLE & & \\
    RKE Symmetric-Key Cyrptography\cite{glocker_defense_2017} & 4 & 4 & & & \CIRCLE & & & \CIRCLE & \CIRCLE &  \\
    General RKE strengthening\cite{patel_defense_2018} & 13 & N/A & & & \CIRCLE & & & \CIRCLE & \CIRCLE &  \\
    Timestamp-based defense\cite{grenne_replay_2020} & N/A & N/A & & & \CIRCLE & & & \CIRCLE & &  \\
    PUF-based RKE mutual authentication\cite{parameswarath_puf_rolljam_2022} & 0.103 & 0.097 & & & \CIRCLE & & &\CIRCLE & &  \\
    Authentication tailoring RollJam \cite{parameswarath_auth_rolljam_2022} & 10 & 10 & & & \CIRCLE & & &\CIRCLE & &  \\
    Quantum safe RKE Authentication \cite{parameswarath_pq_rolljam_2022} & N/A & N/A & & & \CIRCLE & & &\CIRCLE & &  \\
    Distance bonding based RKE\cite{francillon_relay_2011} & N/A & N/A & & \CIRCLE & & &  & \CIRCLE & &  \\
    UWB with Pulse Recording\cite{singh_uwb_2019} & N/A & N/A & & \CIRCLE & & \CIRCLE & & \CIRCLE & \CIRCLE &  \\
    Context-Based Secure RKE\cite{wang_cskes_2019} & N/A & N/A & & \CIRCLE & & & & \CIRCLE & &  \\
    Time-based Countermeasure on PKES\cite{xie_relay_2019} & N/A & N/A & & \CIRCLE & & & & & \CIRCLE &  \\
    Machine Learning relay protection\cite{ahmad_ml_2020} & N/A & N/A & & \CIRCLE & & & & \CIRCLE & &  \\
    HODOR fingerprinting for RKE\cite{hodor_ndss_2020, hodor_analysis_2020} & <500 & <500 & & \CIRCLE & \CIRCLE & & & \CIRCLE & \CIRCLE &  \\
    \hline
    \end{tabular}
\end{table*}

In general, academic research to defend the \ac{rke} systems in automotive is recent and started around 2017.
The sole work preceding this time is coping with the significant threat of power analysis, which was published in 2009~\cite{defense_power_2009}.
The authors proposed a \ac{rke} with \ac{prng} and dynamic re-keying strategy to ensure a unique session key, thwarting side-channel attacks that rely on fixed keys.
To work, the transmitter and the receiver synchronize the \ac{prng}s and initialize with random input and output securely transmitted.
This remote entry is resistant to side-channel analysis and ensures privacy against profiling and the manufacturer, which can not exploit the system with a global key.

\subsection{Replay and RollJam Defenses}
The replay attack has been present since the introduction of the first \ac{rke} systems due to the ease of application, especially on fixed codes and in the absence of timestamps.
In addition, as demonstrated by ~\cite{rolljam}, even rolling codes do not secure entirely against them.
In this case, the attacker needs to be in the range of the real remote system and steal the signals while jamming it when the car's owner closes or opens it.
The requirements are more complicated to achieve than in a relay attack, but it is still feasible and explored in the wild.
The defender side needs to get a more complex authentication mechanism without adding a noticeable delay by the user.
The mechanism could add cryptographic operations or more steps in the authentication, which means a trade-off between security and usability.
In 2017, Glocker et al. tried to end this threat by introducing a symmetric key encryption scheme in the \ac{rke}~\cite{glocker_defense_2017}.
The key fob and the onboard computer authenticate themselves using 2000 randomly generated numbers allocated in memory.
Both parties then compare the received messages by using their memory locations.
The protocol secures \ac{rke} against scan attacks, playback attacks, two-thief attacks, and jamming with a specific implementation built on the car.
A year later, in 2018, researchers briefly analyzed the security of vehicles \ac{rke} systems and proposed a defense mechanism involving a \ac{suc} with a tamper-resistant physical identity module used for key derivation and encryption operations~\cite{patel_defense_2018}.
The security comes from using the unclonable security module and builds its system over vulnerabilities found in ~\cite{glocker_defense_2017}.
They opted for pseudo-identities for the vehicle and encrypted challenge-response messages.
The authors also implemented the proposed system using Raspberry Pi 3 and piVirtualWire library to evaluate its practicality. They found that the computational time of the key fob was acceptable for real-world use.
The timestamp-based defense mechanism proposed in ~\cite{grenne_replay_2020} deals directly against the replay attack, enhancing the existing rolling code \ac{rke} systems.
The proposal adds a timestamp and a second-factor authentication to a randomly generated rolling code, all encrypted with \ac{aes} with a 16-byte key.
The receiver determines the validity of the received signal, checking if the timestamp is in a window of 100 seconds.

Lastly, Parameswarath et al. have proposed three different mechanisms to defend specifically against the RollJam attack.
In the first paper, they presented a \ac{puf}-based lightweight mutual authentication~\cite{parameswarath_puf_rolljam_2022}.
Using \ac{puf}s, the system makes the prediction or replication of the challenge-response messages impractical.
Theoretically, using these primitives makes the protocol faster and more secure than previous proposals, but the authors only formally proved it.
Instead, the second work uses hashing and asymmetric cryptography, but it requires a setup phase to transmit the public key~\cite{parameswarath_auth_rolljam_2022}.
In the authentication phase, the key fob generates a random number, combines it with the current date and time, and calculates the hash value, then transmits it with a signature.
Also, only informal security analyses and simulations are present in this paper.
Lastly, Parameswarath et al. proposed a quantum-safe authentication protocol with \ac{qkd}~\cite{parameswarath_pq_rolljam_2022}.
It aims to build a secure symmetric key against quantum attacks.
The advantage of this system is the possibility of understanding the presence of an eavesdropper.

\subsection{Relay Protection}
The relay attack, introduced in 2011 by ~\cite{francillon_relay_2011}, is still a threat nowadays, with cars stolen with the same technique described in Section~\ref{sec:attacks}.
The relay attack is easy to perform and does not require very costly hardware, and due to this reason, more attackers adopt it to steal cars.
On the opposite, defending against it is challenging due to the physical nature of the attack and the requirement to use different hardware and physical layers to cope with it.
Also, the time frame in which it is possible to carry the relay attack is very short, leading to a more difficult detection.
In ~\cite{francillon_relay_2011}, the authors proposed the first possible high-level solution based on distance bounding protocols.
This method requires accurate distance measurements and multilateration with multiple devices inside the car while trusting the key fob.
The car performs a secure distance bounding protocol and unlocks the doors if the key fob is in a specific range.
Singh et al. proposed the use of \ac{uwb} with \textit{pulse reordering} in 2019 to provide resilience to physical-layer attacks with high performance~\cite{singh_uwb_2019}.
It uses random reorders of \ac{uwb} pulses associated with a set of consecutive bits using a permutation that is a shared secret between the communicating devices.
Additionally, it applies a cryptographic XOR operation to the polarity of pulses with a pseudo-random sequence, making the pulse sequence appear random to an attacker.
This system is compatible with pre-existing \ac{uwb} technologies and standards to provide better interoperability and employability.
The authors provided experimental evaluations to demonstrate that \ac{uwb} with pulse reordering achieves a low bit error rate comparable to standard implementations, proving that security enhancements do not degrade communication performance.

Researchers presented two other works in the same year.
The first paper introduced a context-based \ac{rke} system to prevent relay attacks based on \ac{ble} communication between the key and the vehicle~\cite{wang_cskes_2019}.
Collecting data from the environment such as signal strength indicator, messages round-trip time, \ac{gps} coordinates on the key fob, and a radio environment with Wi-Fi, this method determines the key fob's proximity to the vehicle.
The implementation uses a smartphone as a key fob and detects anomalies (with a machine learning algorithm such as Decision Tree) while preventing relay attacks.
Instead, the second paper examines time-based countermeasures for relay attacks against \ac{pkes} systems~\cite{xie_relay_2019}.
They compared two time-based measurements, \ac{toa} and \ac{dtoa}, in distance-bounding protocols.
The authors presented three different \ac{toa}-based estimation methods after concluding that the \ac{dtoa} is not suitable for distance-bounding applications due to its higher uncertainty.
Simulations are available to support their protocols for estimation accuracy.
The three methods provide a trade-off between accuracy and computational load (accompanied by energy consumption).
The relay protection can also rely on machine learning algorithms, as demonstrated by Ahmad et al.~\cite{ahmad_ml_2020}.
Their solution detects relay attacks using machine learning on key fob signals (time, location, signal strength). 
It verifies the driver via \ac{lstm} neural networks analyzing driving behavior, such as acceleration and braking. 
It prevents unauthorized access and driving with 99.8\% and 81\% accuracy.
The algorithm is implemented on the vehicle side, which will unlock the door or not depending on detecting irregularities, working as an intrusion detection system.

Similarly to this work and extending the fingerprinting technique to the key fob, we have \textit{HODOR}\cite{hodor_ndss_2020}, also along with a separate paper analyzing the solution with a real-world setup~\cite{hodor_analysis_2020}.
The paper presents a new formal attack model that covers both \ac{rke} and \ac{pkes} systems.
In this model, the fingerprinting method aims to identify legitimate key fobs while detecting bad actors. 
HODOR runs on an external device that can capture and analyze packets on \ac{uhf} band without modifying the commercial key fobs already on the market.
The fingerprinting technique extracts the signal features from the pulse preamble: the peak frequency, the carrier frequency offset, the \ac{snr}, and other statistical features.
From the data, HODOR requires a training phase for the classifier that categorizes a signal as if it is from an attacker.
In addition, HODOR can handle different environmental factors that could affect the classification.
The results are promising, but the authors argue that the proposed method is still insufficient for practical use. 

In the literature and other media, we also find suggestions to mitigate the replay and relay attacks, such as in ~\cite{francillon_relay_2011}, where a first short-term suggestion is to shield the key fob to avoid relay.
Drastically, the authors also proposed to remove the battery from the device.
Other solutions advise rolling back to physical locks or deactivating the remote entry (if the vehicle allows it) to reactivate it the next time the driver approaches the car~\cite{pkes_protecting_2025}.

To conclude, the defense side must also cope with the new threats on web and car connectivity. 
In this case, the defender is at a disadvantage due to the high complexity of the systems and the need to secure everything to an attack that only needs one entry point.

%-------------------------------------------------------------------------------
\section{Attacks And Defenses Takeaways}\label{sec:takeways}
%-------------------------------------------------------------------------------
After presenting the different attacks and defenses proposed by security researchers over the years, we analyze the current situation in the automotive scenario regarding \ac{rke} and \ac{pkes} systems.
In this section, we answer the first three questions regarding the changes in the attack types and the overall security of the \ac{rke} (Question 1 and Question 2).
Then, we answer Question 3, analyzing the effectiveness of the defense mechanisms proposed and their deployment in the real world.
However, we can only make an analysis based on public information due to the close source systems of car manufacturers.
Based on the research carried out in this paper, we deduce that
\begin{answerbox}
\textbf{\textit{A1. The \ac{rke} and \ac{pkes} systems are generally more secure thanks to their evolution in technology and techniques.}}
\end{answerbox}
\noindent Nonetheless, the security regards principally the hardening against the cryptographic functions and the difficulty in reverse engineering the newly adopted systems such as \ac{ble} and \ac{nfc}.
Additionally, adopting regulations will positively impact the general security of these devices.
This does not mean breaking them is impossible, as we saw in the attack section.
On the contrary, vulnerabilities in new systems let attackers perform malicious actions more freely without the need for user interaction or noticing.
Furthermore, new technologies, such as web services and \ac{ble}, enlarge the attack surface.
To conclude this answer, the \ac{rke} systems available also now could integrate older technologies (even fixed code) and still be vulnerable to most of the attacks presented in this paper~\cite{ayyappan_rolljam_2024}.
In general, the \ac{rke} systems have better security mechanisms in place to defend against typical cryptographic attacks but still lack security against relay and replay attacks, which are still a threat to these devices.
The physical layer attacks are more complex to defend against, leading the shift to new technologies such as \ac{uwb}.
This brings us directly to answering the second question:
\begin{answerbox}
\textbf{\textit{A2. The attack types did not change during the last 15 years. This is mostly due to the relay attack's exploitability, ease, and diffusion~\cite{relay_hitcon_2017}.}}
\end{answerbox}
\noindent In fact, as anticipated in Section~\ref{sec:implementation}, in legacy technologies, the focus was on cryptographic operations, but the physical layer remained the same.
This allowed the attackers to exploit RollJam and relay attacks in more advanced systems.
In fact, the jamming and replay attack is still a viable option for the thieves.
The hardening in the cryptographic functions made cryptographic attacks harder to perform.
However, the relay is still possible even in \ac{ble} and \ac{nfc} technologies, as we discussed in Section~\ref{sec:attacks}.
Above all, the \ac{uwb} promises of relay protection are in discussion in this last year due to possible attacks against such systems~\cite{tesla_uwb_vicone_2024}.
Moreover, the introduction of web services and \ac{api} brought a new paradigm in the automotive scenario with the need to adapt the defenses and security measures to a broader scope.

Regarding the effectiveness of the defenses, we can infer that it is harder to exploit cryptographic weaknesses or find them in the wild.
However, we can infer that:
\begin{answerbox}
\textbf{\textit{A3. All the different proposals found in the literature are not implemented in real systems, thus reducing the effectiveness of the defense methods developed by the academic research.}}
\end{answerbox}
\noindent The motivation must be found in the developing of schemes that require various hardware and software implementation, with maintenance by the car manufacturers.
Additionally, as we show in Table~\ref{tab:defenses}, in general, these works miss effective and real studies on the possible implementations and computational costs they have on the system.
Only a few of them consider the cost of introducing new protocols.
Still, Joo et al. considered the 500 ms threshold admissible due to the difficulty for a human to notice any difference~\cite{hodor_analysis_2020}.
Also, the development costs for \ac{rke} systems can rise due to the dedicated hardware needed for the proposed systems, which can discourage companies from adopting such methods.
Moreover, different proposed techniques did not have a real-world study case or experimentation, supporting their results based only on simulations, thus not impacting the choices of \ac{oem}s in adopting such countermeasures.
To conclude, most works attempt to contrast the Relay and RollJam attacks.
The different assumptions and hardware setups affect the possible deployability, especially when considering a setup phase that uses a \ac{puf}-based or QKD-based protocols~\cite{parameswarath_puf_rolljam_2022, parameswarath_pq_rolljam_2022} that need special hardware and secure channels with trusted nodes.

%-------------------------------------------------------------------------------
\section{Future Directions}\label{sec:future}
%-------------------------------------------------------------------------------
The research in \ac{rke} and \ac{pkes} system security is gaining ground in the automotive scenario, and it is needed to keep our cars secure from thieves.
The answer to the last research question is as follows:
\begin{answerbox}
\textbf{\textit{A4. Car manufacturers used the security-by-obscurity paradigm to prevent attacks. The change in this mentality will strengthen the \ac{rke} and \ac{pkes} technology, with the cooperation of security researchers and more open specifications and details of such technologies.}}
\end{answerbox}
\noindent
In research and development, the study and adoption of \ac{uwb} is promising but needs more robust implementations to protect against relay attacks. 
However, to achieve this result, we need companies to trust the security community and seek more security auditing and testing, especially now that web vulnerabilities are much easier to introduce in complex systems.
First, the open source paradigm can offer transparency, faster security updates, and an entire community searching for bugs and patching vulnerabilities.
On the contrary, closed-source development, generally adopted by companies, relies only on the black-box approach from the attacker's perspective, but different examples in various environments show how this approach is fallacious.
A good starting point is what we saw last year, and it has already been replicated for 2025 with the introduction of the Pwn2Own Automotive competition, where companies and the best security researchers meet to test the latest technologies~\cite{pwn2own}.
Regarding academic research, the new focus should target the web applications and \ac{api}s, with new methods to precisely tailor the automotive case studies and find possible vulnerabilities and wrong paradigms assumed by the developers.
In addition, \ac{uwb}, Bluetooth, and \ac{nfc} technologies can still be a valuable research direction to stress out due to the technical difficulties in following the specifications without introducing errors.
The use of standards and normative helps in developing new \ac{rke} systems that are secure against known attacks.
In fact, as discussed in Section~\ref{sec:background}, the ISO/SAE 21434 provides guidelines to build a secure system only as requirements and not on how to implement it with specific technical characteristics. 
The presence of a specific standardization and technology to use in creating \ac{rke} system can thus provide better regulations for \ac{oem}s.
At the same time, there is also the need for more agencies to assist at the developing stage and control the conformity to the regulations.

%-------------------------------------------------------------------------------
\section{Related Works}\label{sec:related}
%-------------------------------------------------------------------------------
Only a few works tried to assess the security and point out the current situation about \ac{rke} and \ac{pkes} systems.
Most manuscripts include them in a broader discussion about a comprehensive automotive security review, indicating remote entry as an attack surface~\cite{survey_2011, valasek_survey_2014, jing_survey_2024}. 
Among the \ac{rke} specific works, Tillich and Wójcik presented a study on a particular car immobilizer application by Amtel, implementing an open security protocol stack~\cite{survey_2012}. The authors described the five theoretical attacks they found against it without implementation.
Only two works comprehensively review the technologies utilized in vehicle remote access.
The first one presents a security analysis and replication of simple attacks by Breuls~\cite{breuls_security_2019}.
In the thesis manuscript, the author describes the history of the remote keyless entry, provides a general representation, and provides a specific case of KeeLoq in Microchip Technology.
After that, the author implements the Jamming and Replay attack with a detailed technical description.
The second and most recent, by Zheng et al.~\cite{new_survey}, is similar to this work but only discusses the technologies in a high-level overview, briefly presenting some of the most well-known attacks. 
Furthermore, it does not cover the new technologies and web application interface for remote keyless entry.
Nonetheless, the authors present a good overview of these systems, especially regarding defense mechanisms.

%-------------------------------------------------------------------------------
\section{Conclusions}\label{sec:conclusions}
%-------------------------------------------------------------------------------
This work presented the first comprehensive systematic review of attacks and defenses tailoring remote keyless vehicle entry.
The goal of this systematization of knowledge is to assess the situation nudging the evolution of the \ac{rke} during the years and how attacks have afflicted it that persist still now.
We described the attack classes on different technologies in such systems and how researchers tried to cope with them, presenting various solutions.
We analyzed the problems that brought to the current state, with relay and RollJam as a threat to even the most advanced \ac{rke} systems, and how the web services enlarged the attack surface.
Finally, we hint at the topics researchers can follow to make these systems more secure.

% %-------------------------------------------------------------------------------
% \section*{Acknowledgments}
% %-------------------------------------------------------------------------------

% The USENIX latex style is old and very tired, which is why
% there's no \textbackslash{}acks command for you to use when
% acknowledging. Sorry.

%-------------------------------------------------------------------------------
\Urlmuskip=0mu plus 1mu\relax
\bibliographystyle{plain}
\bibliography{bibliography}

\begin{thebibliography}{10}
\bibitem{why_third_party}
Report: {Why} drivers are turning to third-party car apps · {Smartcar} blog.
\bibitem{bmw_uwb}
What’s the deal with {Ultra} {Wideband} technology and what will it do for your car? {\textbar} {BMW}.com.
\bibitem{stealing_cars}
‘{The} issue is significant’: {Experts} issue keyless car theft warning.
\end{thebibliography}

%%%%%%%%%%%%%%%%%%%%%%%%%%%%%%%%%%%%%%%%%%%%%%%%%%%%%%%%%%%%%%%%%%%%%%%%%%%%%%%%
\end{document}